\documentclass[12pt]{reportj}
\usepackage{natbib}
\setcitestyle{aysep={}} 
\renewcommand{\bibname}{}

\usepackage{deluxetablej}
\usepackage{hyperref} 
\makeatletter
\def\@to{to}
\makeatother
\usepackage{times}
\usepackage{graphicx}
\usepackage{tocloft}
\usepackage{fancyheadings}
\usepackage{xcolor}
\usepackage{booktabs}
\usepackage{longtable}

\emergencystretch=\maxdimen
\usepackage{datetime}
\newdateformat{ddmonthyyyy}{\THEDAY\ \monthname[\THEMONTH]\ \THEYEAR}

\renewcommand\thesection{\arabic{section}}
\renewcommand\thesubsection{\thesection.\arabic{subsection}}

\def\ssection#1{\setcounter{subsection}{0} \refstepcounter{section} \section*{\hbox to \hsize{\large\bf \arabic{section}. #1\hfill }}\label{sec} \addcontentsline{toc}{section}{\arabic{section}. #1}}
\def\ssubsection#1{\setcounter{subsubsection}{0} \refstepcounter{subsection}\subsection*{\hbox to \hsize{\normalsize\bfseries\itshape \arabic{section}.\arabic{subsection} #1\hfill}}\label{subsec} \addcontentsline{toc}{subsection}{\arabic{section}.\arabic{subsection} #1}}
\def\ssubsubsection#1{\refstepcounter{subsubsection}\subsection*{\hbox to \hsize{\normalsize\it \arabic{section}.\arabic{subsection}.\arabic{subsubsection} #1\hfill}}\label{subsubsec} \addcontentsline{toc}{subsubsection}{\arabic{section}.\arabic{subsection}.\arabic{subsubsection} #1}}

\def\ssectionstar#1{\section*{\hbox to \hsize{\large\bf #1\hfill}} \addcontentsline{toc}{section}{#1}}
\def\ssubsectionstar#1{\subsection*{\hbox to \hsize{\normalsize\bfseries\itshape #1\hfill}} \addcontentsline{toc}{subsection}{#1}}
\def\ssubsubsectionstar#1{\subsection*{\hbox to \hsize{\normalsize\it  #1\hfill}} \addcontentsline{toc}{subsection}{#1}}

\renewcommand{\cftaftertoctitle}{%
\mbox{}\hfill{\normalfont Page}}
\begin{document}

~\\

\vspace{-2.4cm}
\noindent\includegraphics*[width=0.295\linewidth]{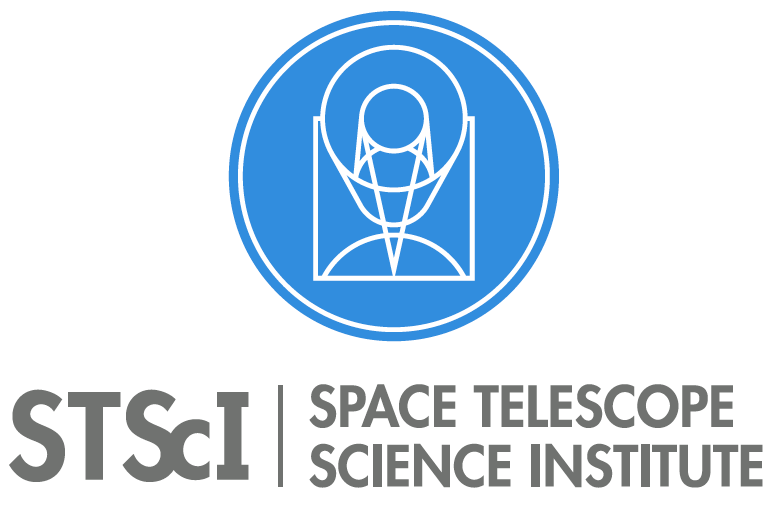}

\vspace{-0.4cm}

\begin{flushright}
{\bf Instrument Science Report COS 2026-06(v1)}
    
\vspace{1.1cm}
    
{\bf\Huge The Long-Term Radial Velocity Stability of the COS G130M/1291 Mode}
    
\rule{0.25\linewidth}{0.5pt}
    
\vspace{0.5cm}
    
John Debes$^1$, David M. French$^1$, Leonardo A. Dos Santos$^1$, Travis Fischer$^1$, Svea Hernandez$^1$, and Ravi Sankrit$^1$
\linebreak
\newline
\footnotesize{$^1$ Space Telescope Science Institute, Baltimore, MD}
 
\vspace{0.5cm}
    
17 August 2026
\end{flushright}

\vspace{0.1cm}

\noindent\rule{\linewidth}{1.0pt}
\noindent{\bf A{\footnotesize BSTRACT}}

We investigate the long-term radial velocity (RV) stability of one of the Far-UV (FUV) modes on the Cosmic Origins Spectrograph (COS). We do so by using a standard white dwarf, WD~0308-565, that has been observed over a large fraction of COS's operational lifetime of 17 years. We demonstrate that for modest signal-to-noise COS can return an RV stability of $<$3~km~s$^{-1}$ over a baseline of 15 years. For higher SNR and a smaller number of epochs, we obtain $\sigma_{\mathrm V}=600$~m~s$^{-1}$. We also show that the absolute radial velocity accuracy for this target is consistent with results from ground-based optical spectroscopy. We discuss some observational strategies and astrophysical objects that may be well suited to pushing to the very RV limit of the instrument. Utilizing COS for precision RV work opens a new scientific avenue for space-based UV astronomy as COS approaches 20 years of operations and hints at what may be possible with the stability of a high-resolution UV spectrograph on future missions such as the Habitable Worlds Observatory.

\vspace{-0.1cm}
\noindent\rule{\linewidth}{1.0pt}

\newpage

\renewcommand{\cftaftertoctitle}{\thispagestyle{fancy}}
\tableofcontents

\lhead{}
\rhead{}
\cfoot{\rm {\hspace{-1.9cm} Instrument Science Report COS 2026-06(v1) Page \thepage}}


\vspace{-0.3cm}
\ssection{Introduction}\label{sec:intro}
\vspace{0.25in}

The Cosmic Origins Spectrograph (COS) Instrument was installed on May 16, 2009, and is thus entering its seventeenth year of operations. As the Hubble Space Telescope Observatory looks toward its fourth decade of operations, each instrument is being considered for unique uses that cannot be accomplished by other facilities. One such avenue is the very long time baselines that can be achieved in the UV with COS, STIS, WFC3/UVIS, and ACS/SBC imaging and spectroscopy. Long time baseline spectroscopy that is accurate in both flux and wavelength is useful for many applications, such as the evolution of nearby stars or the characterization of binary orbits with UV bright companions.

Historically, COS far-ultraviolet (FUV) modes have absolute and relative radial velocity (RV) accuracy that typically precludes sub-km/s precision. The multiple changes in lifetime position (LP) on the detector raises the possibility that subtle systematic offsets in absolute RV accuracy could arise. The geometric distortion of COS \citep{indriolo25, kakkad25,indriolo25b} and detector walk \citep{hasselquist25,hasselquist25b} has also limited RV accuracy across the FUV modes. The recent upgrade to the FUV geometric distortion correction and the detector walk has produced the highest quality wavelength solutions across the entire operational lifetime of the COS FUV detector, promising relative accuracies for medium resolution modes of $<$3~km~s$^{-1}$ (cf., \href{https://hst-docs.stsci.edu/cosihb/chapter-1-an-introduction-to-cos/1-3-cos-quick-reference-guide}{COS Instrument Handbook}).  Epoch-to-epoch accuracy is dominated by the stability of the wavelength zero points, which for G130M/1291 are at the level of $\sim$1 COS pixel \citep[3~km~s$^{-1}$][]{fischer21,french24}. For the COS FUV, $\sim$6 pixels span a spectral resolution element (resel) across most LPs.

We asked the exploratory question: can COS deliver this stability or better across its 17 year lifetime? We demonstrate in the following report that, in the case of an astrophysical object with sufficient continuum flux,  intrinsic absorption lines, and unsaturated or mildly saturated interstellar medium (ISM) absorption features, the answer is yes.  Using the white dwarf (WD) standard star WD~0308-565 in combination with two carbon photospheric lines, we obtain a stability of $\sim$2.6~km~s$^{-1}$ over the lifetime of the COS instrument for the G130M/1291 mode for spectra with a median signal-to-noise per resel (SNR) of 18. We also show that better precision may be possible with other targets or techniques and that the COS data rule out stellar and substellar companions across a range of semi-major axis for WD~0308-565 for the first time. 

In \S\ref{sec:data} we discuss our use of HASP products to facilitate our collection of high-quality, relevant datasets for our investigation. In \S\ref{sec:method} we detail how we measured the RV of WD~0308-565 from archival calibration and monitoring data, and in \S\ref{sec:results} we demonstrate COS's radial velocity stability in the absence or presence of ISM reference lines. Finally, in \S\ref{sec:conc} we discuss other steps that may allow users to push below the 2-3~km~s$^{-1}$ limit.


\vspace{-0.3cm}
\ssection{Datasets Used}\label{sec:data}

COS has routinely observed the nearby WD~0308-565 for flux calibration, measurement of the G130M spatial profile, and monitoring of the COS time-dependent sensitivity evolution in the FUV since 2011. Some of these observations were taken at sub-optimal exposure times or modes. 

Datasets marked for HASP coaddition in the MAST archive are filtered to remove any observation that may have been a non-standard calibration observation or suffered some observatory anomaly. We therefore selected the subset of archival WD~0308-565 datasets that was used in the HASP products and that had sufficient exposure time to have SNR$>$14. 

We downloaded the data using the \href{https://mast.stsci.edu/search/ui/#/hst}{HST Mission Search Form}, selecting the G130M grating, 1291 central wavelength (CENWAVE), t$_{exp}>100$~s, and HASP=1 (\href{https://mast.stsci.edu/search/ui/#/hst/results?resolve=true&target=WD+0308-565&radius=3&radius_units=arcminutes&data_type=spectrum&observations=S&active_instruments=cos&exposure_time=%3E=100&spectral_element=G130M&custom_col_sel_0=sci_central_wavelength&custom_col_val_0=1291&custom_col_sel_1=sci_haspnum&custom_col_val_1=1&select_cols=sci_central_wavelength,sci_data_set_name,sci_dec,sci_actual_duration,sci_spec_1234,sci_haspnum,sci_instrume,sci_pi_last_name,sci_pep_id,sci_ra,sci_start_time,sci_stop_time,sci_targname&search_key=e05383d47d4ce&useStore=false}{Search link here}). The observations covered the time period from 7 January 2011 to 27 January 2026, for a total baseline of fifteen years. The median exposure time was 267~s, the minimum exposure time was 208~s, and the maximum exposure time was 3160~s.

\begin{figure}
    \centering
    \includegraphics[width=\linewidth]{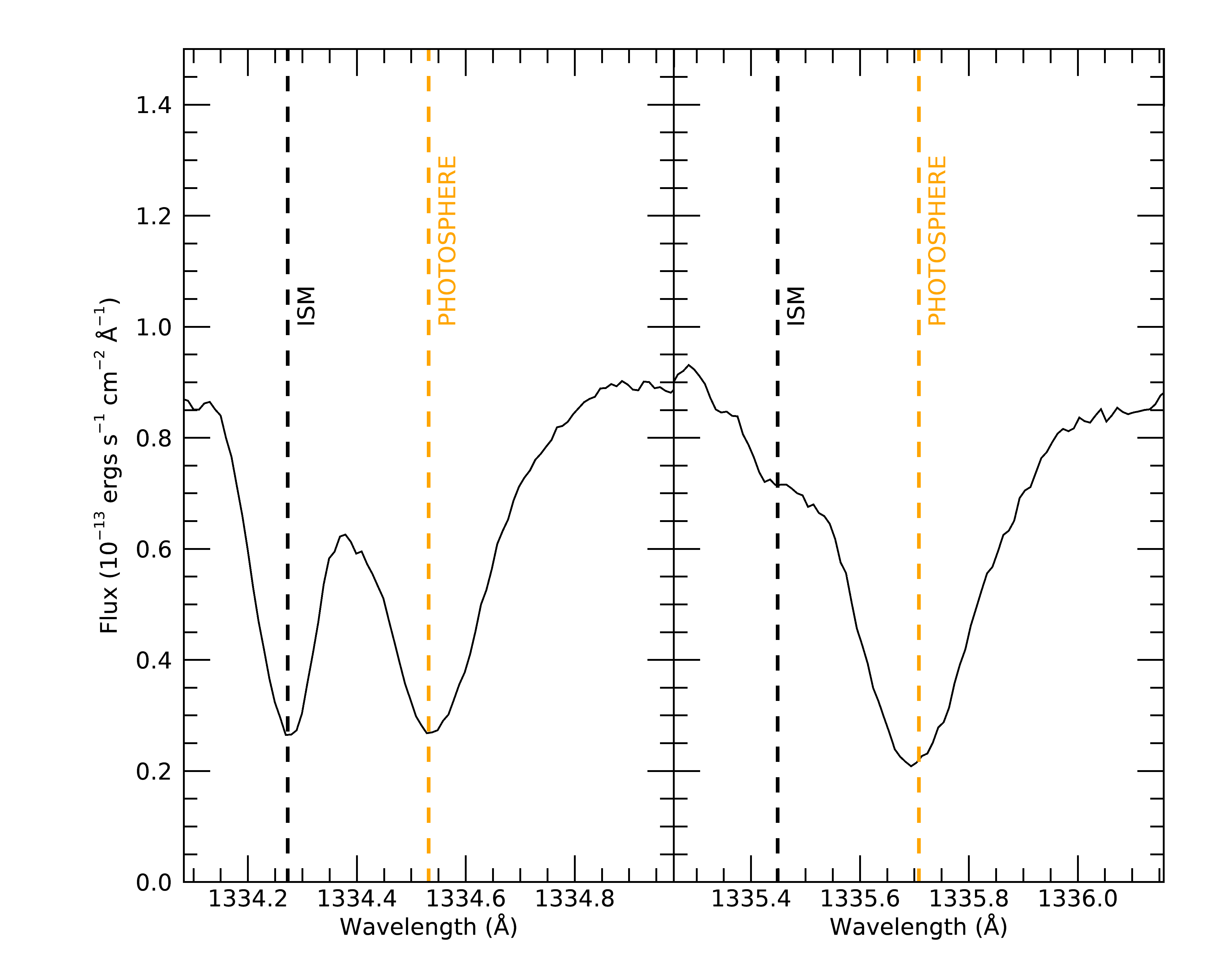}
    \caption{High SNR spectrum of WD~0308-565 demonstrating the three spectral lines we use for the measurement of radial velocities. The left spectrum shows the photospheric C~II 1334~\AA\ line (at V $\sim68$~km~s$^{-1}$) and the C~II ISM line to the blue of the photospheric line (at V $\sim10$~km~s$^{-1}$). The right spectrum shows the strong photospheric line for C~II at 1333~\AA\ and a possible weak ISM line. The weak ISM line is not used in this analysis.}
    \label{fig:spec}
\end{figure}

\vspace{-0.3cm}
\ssection{Methodology}\label{sec:method}

There are a number of ways to measure the radial velocity of an object. WD~0308-565 is a hot DBQ white dwarf whose UV spectrum exhibits several ISM lines as well as photospheric carbon \citep{petitclerc05}. The strongest photospheric lines are the C~II lines at 1334.532 and 1335.708~\AA. Fortuitously, there also exists a strong C~II 1334 ISM line that is blue shifted from the photospheric line due to the motion of the WD with respect to the local intervening ISM. Figure \ref{fig:spec} shows the spectral regions of interest. WD~0308-565 has a radial velocity of $\sim$68~km~s$^{-1}$, while the ISM line is at $\sim$10~km~s$^{-1}$. For the median exposure, a single FP-POS was obtained for 267~s. These observations were part of the time-dependent sensitivity monitoring programs. Since the G130M grating is losing sensitivity over time, the exposure time of individual exposures has been increased to maintain a roughly consistent SNR/resel, making these observations relatively homogenous near the C~II lines: 119 datasets had SNR=14-20, while 20 datasets had an SNR exceeding 20. The median SNR was 18.

We employed a Voigt fitting routine to measure the center of the two photospheric lines. We fit portions of the spectrum in a window $\pm$55~km~s$^{-1}$ in velocity space to fit the line core. We found that Voigt fitting was not as successful for the ISM line due to blending with the stronger photospheric line and instead used flux weighted centroiding. For the ISM line we calculated the centroid in a window of $\pm$25~km~s$^{-1}$ from the rough center of the ISM line in velocity space. 

There are other methods to determine the RV of each line that may have less systematic uncertainties, such as due to fitting errors. In particular, the COS line spread functions (LSFs) are not strictly Gaussian or well-described by a Voigt profile.  Further, the line profile changes with different LP moves. Cross-correlation of the spectra with a model template convolved by LP-dependent (LSFs) could also be employed to measure RVs and may be considered for future investigations of RV with COS.

\vspace{-0.3cm}
\ssection{Results}\label{sec:results}

In this section we describe our results, considering first the full data set and then a subset of high-SNR observations. The latter demonstrate the potential for sub-km~s$^{-1}$ RV precision in the case of an object with many lines.

\subsection{RVs from All Datasets}
\begin{figure}[b!]
    \centering
    \includegraphics[width=\linewidth]{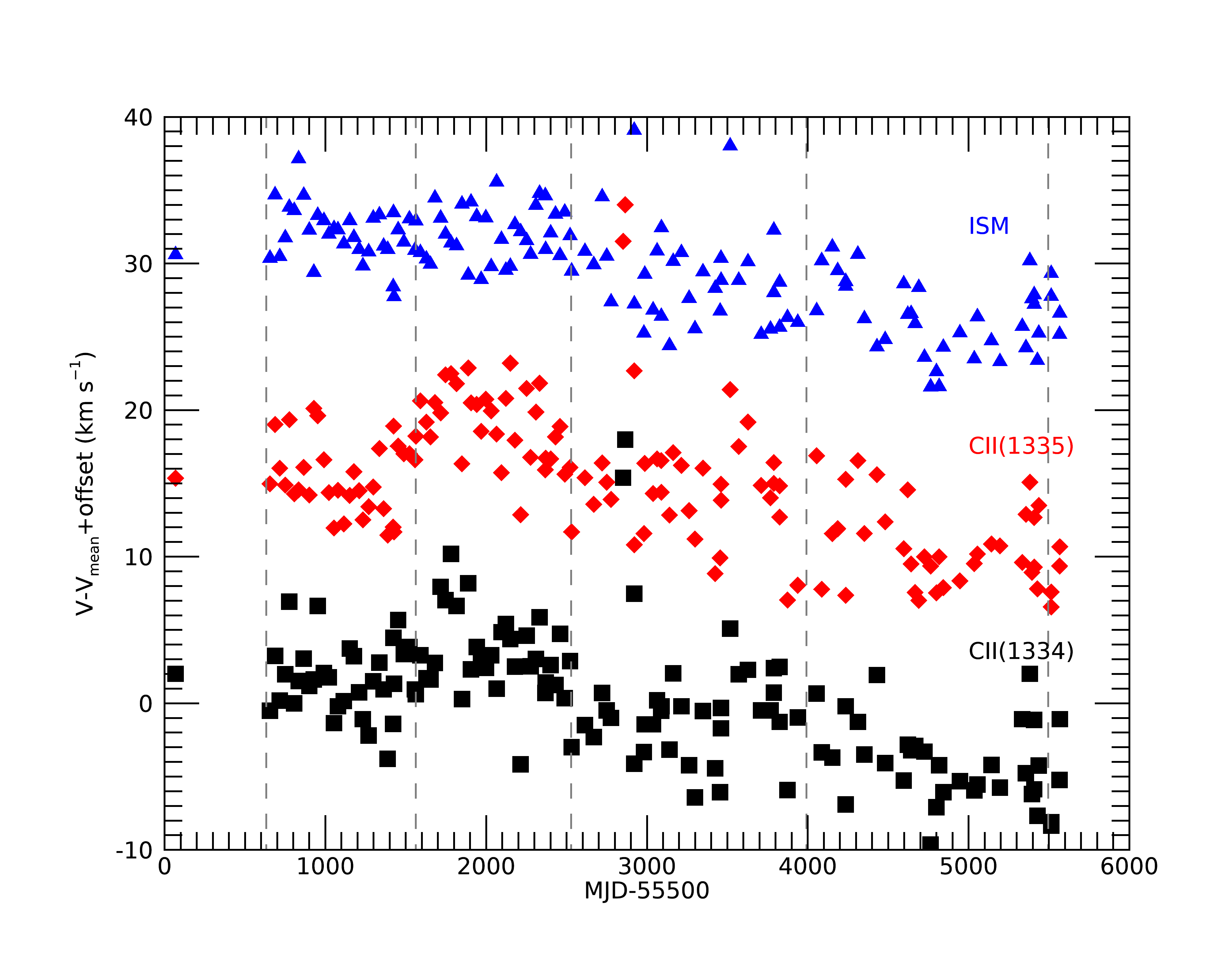}
    \caption{RV measurements of three C~II lines observed in the spectrum of WD~0308-565. The blue triangles are from the ISM line, red triangles are from C~II (1335), and the black squares are from C~II (1334). A noticeable trend toward negative velocities is present with time in all three lines, indicative of an instrumental or observatory effect. Grey dashed lines indicate LP changes.}
    \label{fig:rvraw}
\end{figure}

The measured radial velocities for each line (after we subtract the mean velocity and add a constant offset) are presented in Figure \ref{fig:rvraw}. All three velocity curves show a linearly decreasing trend that changes by $\sim$10~km~s$^{-1}$ from the first observation to the last. Since the ISM line should have a constant radial velocity, we interpret this as a systematic instrumental trend. The standard deviation of all velocities for photospheric C~II (1334) and C~II (1335) is 4.0~km~s$^{-1}$, while for the ISM the standard deviation is 3.3~km~s$^{-1}$, consistent with the relative accuracy claims of the COS Instrument Handbook, despite the overall observed trend. The mean velocities of the two C~II lines are 68 and 67~km~s$^{-1}$. The mean velocity of the ISM line is 11~km~s$^{-1}$.

The cause of the small drift is beyond the scope of this report, but is on the order of 600~m~s$^{-1}$/yr or less than 0.3 pixels/yr. This could be caused by drifts in the wavecal lamp, the Optics Select Mechanism 1, or some other optical element that is not obviously dependent on changes in lifetime positions. If it is a wavelength zero-point issue, it should be present in a re-analysis of programs that monitor the wavelength solution of COS over time.

For the purposes of precision RVs, we can use the relative velocity between the ISM line and the photospheric lines to improve the overall standard deviation of the RV measurements, provided the velocities of the three lines are as well correlated as they appear. To quantify the correlation, we calculated a Pearson correlation coefficient of 0.86 for the velocities of the two photospheric lines. We find Pearson correlation coefficients of 0.63 and 0.62 for C~II (1334) and C~II (1335) compared with the ISM line velocities. 

\begin{figure}
    \centering
    \includegraphics[width=\linewidth]{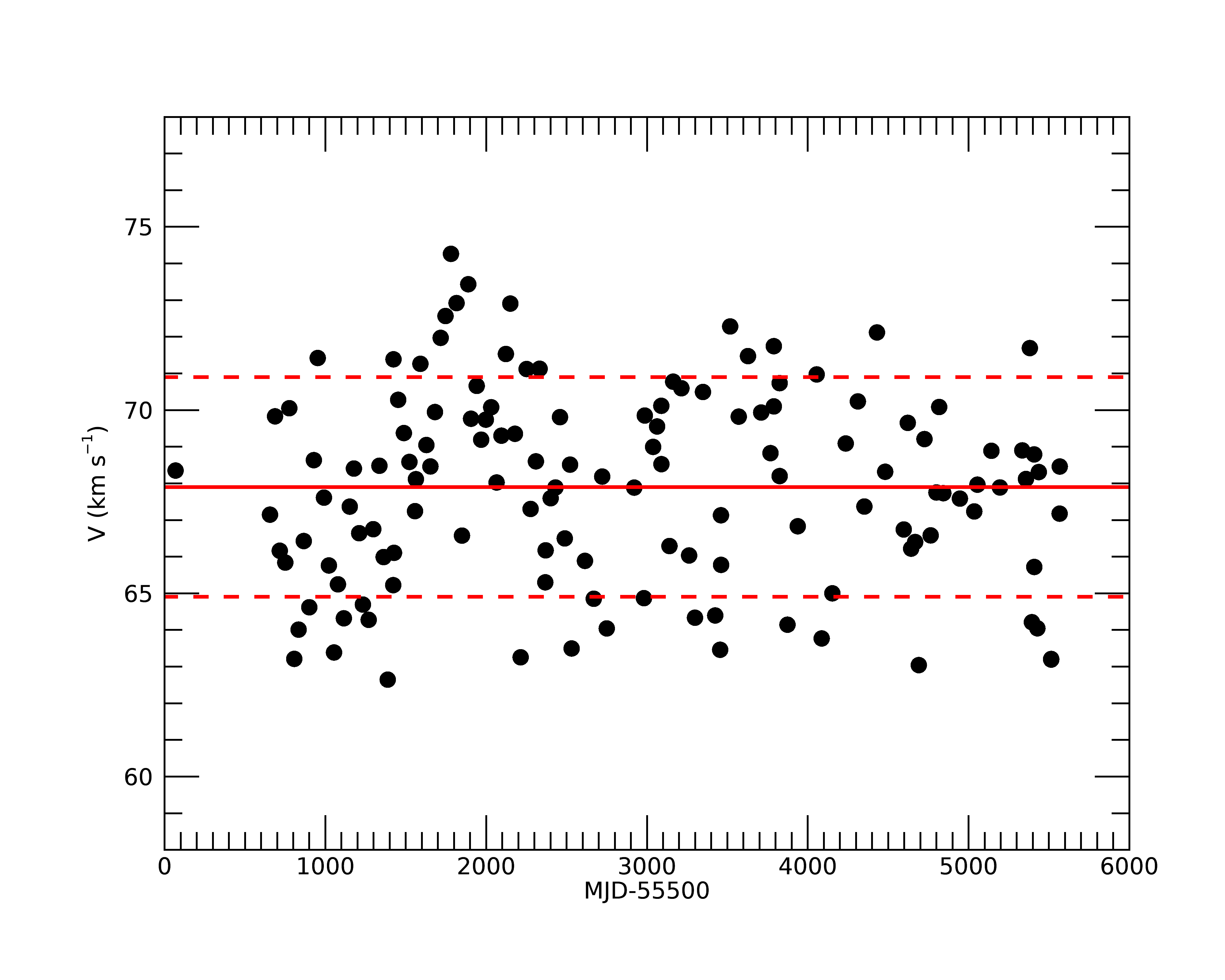}
    \caption{Final RV of WD~0308-565 from the combined and corrected C~II photospheric lines. The red solid line is the mean velocity 68~km~s$^{-1}$, and the dashed lines show $\pm$3~km~s$^{-1}$ accuracy.}
    \label{fig:rvfinal}
\end{figure}

We smoothed the ISM RVs with a boxcar function over 5 samples and subtracted the median ISM velocity to model the systematic trend. We then detrended the photospheric line RVs and averaged the RVs of the two lines. The final RV curve as a function of time is shown in Figure \ref{fig:rvfinal}. We note that the mean velocities of the two C~II lines differ by 1.7~km~s$^{-1}$, consistent with local deviations of the wavelength solution of $<$1 pixel \citep{indriolo25}.

To remove discrepant velocities, we calculated a standard deviation for the sample, removed outliers that were larger than 2.5$\sigma$, and re-calculating the standard deviation of the sample. We repeated this process twice before no further outliers were found. Five corrected velocities from MJDs 58276, 58351, 58364, 58421, 59686, and 59736 were removed from the sample for having anomalously discrepant velocities, with $|\delta~\mathrm{V}|>$7~km~s$^{-1}$.

We find a standard deviation for the velocities of 2.6~km~s$^{-1}$, in line with the 3~km~s$^{-1}$ relative accuracy quoted by the COS Instrument Handbook. The final mean velocity that we measure for WD~0308-565 is 67.9~km~s$^{-1}$.

WD~0308-565 possesses several optical photospheric lines and was observed with the VLT/UVES spectrograph in 2002 \citep{voss07}. The advantage of a UVES observation is that it pins the absolute RV accuracy to better than 500~m~s$^{-1}$ \citep{sacco14}. We obtained publicly available UVES spectra of WD~0308-565 from the ESO archive (Program 167.D-0407, PI: Napiwotzki) and measured the radial velocity of the He~I (5876) line using a Gaussian fit \citep{falcon12}. We obtain a velocity of 64$\pm$7~km~s$^{-1}$, consistent with the COS value. 

\subsection{A High SNR Subset of the WD~0308-565 Dataset}

Our sample includes eleven datasets that span nearly the same temporal baseline as the full sample but have 5-10 times the average exposure time, resulting in SNR$>$30 for each spectrum, allowing us to investigate the RV stability of COS in the case of high SNR data obtained with multiple FP-POS.

We corrected this subset of velocities in the same manner as before and discovered one dataset (LFKD56010) was significantly discrepant. Given that these were early LP7 observations that used split wavecals \citep{dossantos25}, it is possible that there was a particularly large drift in this dataset that was not completely accounted for in the wavecal exposures. We therefore removed it from the sample, leaving ten observations. Of these, we find that the mean velocity is 67.5~km~s$^{-1}$, consistent with the larger sample, and that $\sigma_{\mathrm V}=600$~m~s$^{-1}$. Each individual line had $\sigma_{\mathrm V}=900$~m~s$^{-1}$. As with the larger sample, we note that the mean velocities of the two C~II lines differ by 1.9~km~s$^{-1}$. 

This significant improvement with higher SNR suggests that RV precision could be improved in situations where a target has many lines (both ISM and/or photospheric) in the G130M bandpass. Since these observations meet or exceed the maximum allowable SNR for G130M \citep{romanduval23}, we expect that for COS the single line RV precision will be limited to $\sim$900~m~s$^{-1}$, but the possibility of achieving better precision exists. 

\subsection{Implications for Possible Companions to WD~0308-565}

WDs are compact objects and can possess companions that are not easily detected, such as a companion WD substellar object. WD~0308-565 does not have any nearby co-moving objects detected in Gaia DR3, and its Root unit weight error (RUWE) is consistent with no significant astrometric signatures of a companion. Lower mass unresolved companions, such as brown dwarfs might result in WD~0308-565 possessing a mid-IR excess. However, its K-W2 $\sim -0.01$ color is consistent with no excess, ruling out some substellar companions. To date no RV curve of this WD has been published.

The full G130M/1291 dataset is ideal for searching for the evidence of binary companions: with over 100 observations, the median spacing between epochs is roughly 31 days, and there is a long time baseline, allowing for decent sampling of short periods up to $\sim$15~years. We calculated a Lomb-Scargle periodogram of the velocities.  No significant periodicity was detected with power above the estimated 1\% false-alarm probability (FAP).

\begin{figure}
    \centering
    \includegraphics[width=\linewidth]{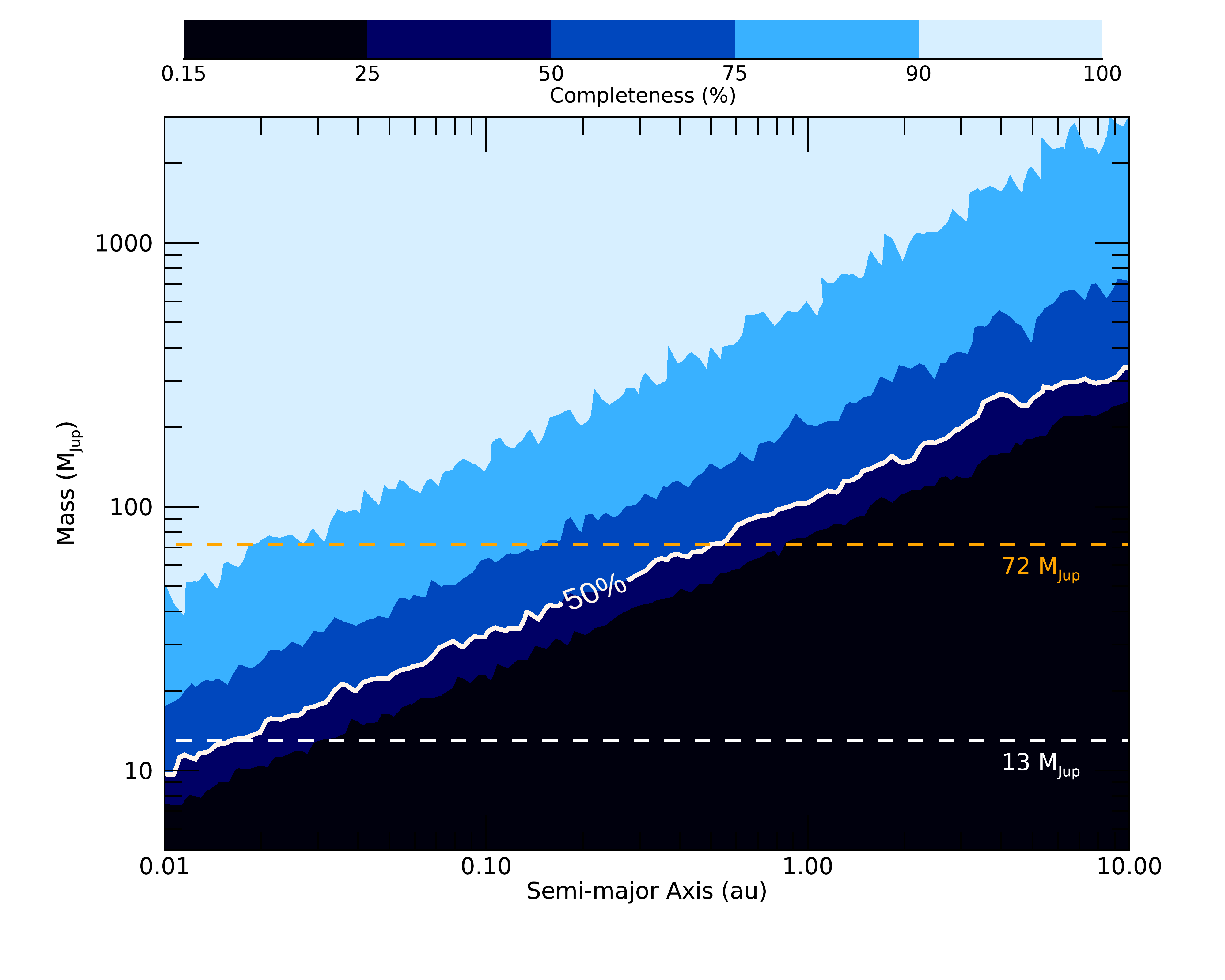}
    \caption{RV Completeness plot for WD~0308-565 as a function of companion mass and semi-major axis. The 50\% completeness contour is highlighted with a white line. Short period stellar and substellar companions are ruled out due to the long RV baseline.}
    \label{fig:complete}
\end{figure}

While no signal is detected, we wish to understand our sensitivity to different types of companion masses and orbits. To that end, we estimated the probability of detecting RV signals from companions with a range of masses and orbital periods. For each mass and period pair, we calculated the possible RV signal from 1000 randomly inclined circular orbits with random initial orbital phases. We injected the artificial signal into our WD~0308-565 dataset and then calculated a $\chi_\nu^2$ value for an assumed flat RV curve assuming our uncertainty per velocity was equivalent to our velocity standard deviation or 2.6~km~s$^{-1}$. If the $\chi_\nu^2$ probability for no variability with the degrees of freedom (133) was $<$10$^{-3}$, we considered the mass and semi-major axis companion detected. We verified that the 50\% completeness contour was similar whether we used this criterion or a significant peak in a Lomb-Scargle periodogram. The completeness of our dataset to signals as a function of mass and semi-major axis is plotted in Figure \ref{fig:complete}

\vspace{-0.3cm}
\ssection{Conclusions}\label{sec:conc}

We have demonstrated that the most popular FUV mode of COS has a long-term RV stability that meets or exceeds the current reported relative wavelength solution accuracy of 3~km~s$^{-1}$ and obtains sub-km~s$^{-1}$ accuracy for a single line at high SNR.  Greater RV precision could be possible for objects with large numbers of absorption features in their spectra. We also demonstrate that WD~0308-565 does not possess any stellar companions within $\approx$0.6~au, and any brown dwarfs within 0.02~au. It appears to be acceptable to use as an RV standard for km~s$^{-1}$ accuracy with COS.

This demonstration of RV precision opens the possibility for long-term RV monitoring of UV-bright objects for a variety of science cases by revisiting past COS targets or by intensively monitoring new objects. A similar RV stability experiment could be conducted on the STIS instrument, which has the capability of higher spectral resolution than COS (R$\sim$40,000--100,000) and a longer operational baseline (1997--2004; 2009--present). However, the Hubble Space Telescope Observatory undergoes significant thermal variations throughout its orbit, which could limit the ultimate RV precision of COS or STIS in the UV. Thermal variations for future UV-capable observatories, such as the proposed Habitable Worlds Observer, will likely be smaller since they will not be in near-Earth orbits, resulting in better RV stability.

We also demonstrated that the absolute wavelength accuracy of COS appears to be within $\sim$3~km~s$^{-1}$ when compared to the UVES instrument, but tests of this absolute accuracy are limited by uncertainties in measurements in the visible. He and H lines for white dwarfs are pressure broadened and thus are not ideal for obtaining high-precision radial velocities. 

An absolute measurement of radial velocity accuracy for the COS instrument could be improved by investigating a large number of WDs with Si~II (1265) absorption. WDs that are actively accreting rocky material from dust often show strong absorption lines of both Si~II (1265) and Ca~II (3933), providing a link between COS and other ground-based spectrographs such as VLT/UVES and Keck/HIRES. Metal lines suffer less pressure broadening and thus allow more-precise RV measurements. Dozens of such targets have already been observed by COS.



\vspace{-0.3cm}
\ssectionstar{Change History for COS ISR 2026-06}\label{sec:history}
\vspace{-0.3cm}
Version 1: 17 August 2026 - Original Document

\vspace{-0.3cm}
\ssectionstar{References}\label{sec:references}
\vspace{-1.0cm}
\bibliographystyle{aasjournal}
\bibliography{references}{}

@ARTICLE{falcon12,
       author = {{Falcon}, Ross E. and {Winget}, D.~E. and {Montgomery}, M.~H. and {Williams}, Kurtis A.},
        title = "{A Gravitational Redshift Determination of the Mean Mass of White Dwarfs: DBA and DB Stars}",
      journal = {\apj},
         year = 2012,
        month = oct,
       volume = {757},
       number = {2},
          eid = {116},
        pages = {116},
          doi = {10.1088/0004-637X/757/2/116},
archivePrefix = {arXiv},
       eprint = {1208.2009},
 primaryClass = {astro-ph.SR},
       adsurl = {https://ui.adsabs.harvard.edu/abs/2012ApJ...757..116F}
}

@MISC{romanduval23,
       author = {{Roman-Duval}, Julia and {Johnson}, Christian I. and {Hasselquist}, Sten and {James}, Bethan and {Rafelski}, Marc and {Sahnow}, David},
        title = "{Mitigation of the Fixed-Pattern Noise on COS: Effective Signal-to-Noise versus Number of FP-POS Spectra Dithers}",
 howpublished = {Instrument Science Report COS 2023-11, 22 pages},
         year = 2023,
        month = may,
        pages = {11},
       adsurl = {https://ui.adsabs.harvard.edu/abs/2023cos..rept...11R}
}

@MISC{dossantos25,
       author = {{Dos Santos}, Leonardo A. and {Gomez}, Sierra R. and {Rowlands}, Kate},
        title = "{Updated parameters for the simulated 600-s wavelength calibration lamp flash}",
 howpublished = {Instrument Science Report COS 2025-13, 10 pages},
         year = 2025,
        month = aug,
        pages = {13},
       adsurl = {https://ui.adsabs.harvard.edu/abs/2025cos..rept...13D}
}

@MISC{hasselquist25b,
       author = {{Hasselquist}, Sten and {Ake}, Thomas and {French}, David and {Indriolo}, Nick and {Sahnow}, David},
        title = "{Determining Y-Walk Corrections for the COS FUV Detector}",
 howpublished = {Instrument Science Report COS 2025-11, 15 pages},
         year = 2025,
        month = jul,
        pages = {11},
       adsurl = {https://ui.adsabs.harvard.edu/abs/2025cos..rept...11H}
}

@MISC{hasselquist25,
       author = {{Hasselquist}, Sten and {Ake}, Thomas and {Dashtamirova}, Dzhuliya and {Debes}, John and {De Rosa}, Gisella and {French}, David and {Kakkad}, Darshan and {Indriolo}, Nick and {Plesha}, Rachel and {Sahnow}, David},
        title = "{Determining X-Walk Corrections for the COS FUV Detector}",
 howpublished = {Instrument Science Report COS 2025-10, 20 pages},
         year = 2025,
        month = jul,
        pages = {10},
       adsurl = {https://ui.adsabs.harvard.edu/abs/2025cos..rept...10H}
}

@MISC{kakkad25,
       author = {{Kakkad}, Darshan and {Indriolo}, Nick and {Ake}, Tom and {Debes}, John and {French}, David and {Hasselquist}, Sten and {Sahnow}, David and {Proffitt}, Charles and {Roman-Duval}, Julia},
        title = "{A Revised Geometric Distortion Correction for the Far-Ultraviolet Detector of the Cosmic Origins Spectrograph}",
 howpublished = {Instrument Science Report COS 2025-09, 21 pages},
         year = 2025,
        month = jul,
        pages = {9},
       adsurl = {https://ui.adsabs.harvard.edu/abs/2025cos..rept....9K}
}

@MISC{indriolo25b,
       author = {{Indriolo}, Nick and {Kakkad}, Darshan and {Ake}, Tom and {Debes}, John and {French}, David and {Hasselquist}, Sten and {Sahnow}, David},
        title = "{Measurement and Implementation of a Delta-Geometric Correction for The COS FUV Detector}",
 howpublished = {Instrument Science Report COS 2025-08, 13 pages},
         year = 2025,
        month = jul,
        pages = {8},
       adsurl = {https://ui.adsabs.harvard.edu/abs/2025cos..rept....8I}
}

@MISC{indriolo25,
       author = {{Indriolo}, Nick and {Ake}, T. and {Debes}, J. and {French}, D. and {Hasselquist}, S. and {Kakkad}, D. and {Rafelski}, M. and {Sahnow}, D. and {De Rosa}, G. and {Fischer}, W. and {Frazer}, E. and {Plesha}, R. and {Proffitt}, C. and {Roman-Duval}, J.},
        title = "{An Overview of Improvements to the COS FUV Geometric Distortion and Walk Corrections}",
 howpublished = {Instrument Science Report COS 2025-07, 17 pages},
         year = 2025,
        month = jul,
        pages = {7},
       adsurl = {https://ui.adsabs.harvard.edu/abs/2025cos..rept....7I}
}

@MISC{french24,
       author = {{French}, David M.},
        title = "{Cycle 29 COS FUV Wavelength Scale Monitor}",
 howpublished = {Instrument Science Report COS 2024-02, 9 pages},
         year = 2024,
        month = feb,
        pages = {2},
       adsurl = {https://ui.adsabs.harvard.edu/abs/2024cos..rept....2F}
}

@MISC{fischer21,
       author = {{Fischer}, William J.},
        title = "{Cycle 27 COS FUV Wavelength Scale Monitor}",
 howpublished = {Instrument Science Report COS 2021-10, 8 pages},
         year = 2021,
        month = dec,
        pages = {10},
       adsurl = {https://ui.adsabs.harvard.edu/abs/2021cos..rept...10F}
}

@ARTICLE{sacco14,
       author = {{Sacco}, G.~G. and {Morbidelli}, L. and {Franciosini}, E. and {Maiorca}, E. and {Randich}, S. and {Modigliani}, A. and {Gilmore}, G. and {Asplund}, M. and {Binney}, J. and {Bonifacio}, P. and {Drew}, J. and {Feltzing}, S. and {Ferguson}, A. and {Jeffries}, R. and {Micela}, G. and {Negueruela}, I. and {Prusti}, T. and {Rix}, H.-W. and {Vallenari}, A. and {Alfaro}, E. and {Allende Prieto}, C. and {Babusiaux}, C. and {Bensby}, T. and {Blomme}, R. and {Bragaglia}, A. and {Flaccomio}, E. and {Francois}, P. and {Hambly}, N. and {Irwin}, M. and {Koposov}, S. and {Korn}, A. and {Lanzafame}, A. and {Pancino}, E. and {Recio-Blanco}, A. and {Smiljanic}, R. and {Van Eck}, S. and {Walton}, N. and {Bergemann}, M. and {Costado}, M.~T. and {de Laverny}, P. and {Heiter}, U. and {Hill}, V. and {Hourihane}, A. and {Jackson}, R. and {Jofre}, P. and {Lewis}, J. and {Lind}, K. and {Lardo}, C. and {Magrini}, L. and {Masseron}, T. and {Prisinzano}, L. and {Worley}, C.},
        title = "{The Gaia-ESO Survey: processing FLAMES-UVES spectra}",
      journal = {\aap},
         year = 2014,
        month = may,
       volume = {565},
          eid = {A113},
        pages = {A113},
          doi = {10.1051/0004-6361/201423619},
archivePrefix = {arXiv},
       eprint = {1403.4865},
 primaryClass = {astro-ph.IM},
       adsurl = {https://ui.adsabs.harvard.edu/abs/2014A&A...565A.113S}
}

@ARTICLE{voss07,
       author = {{Voss}, B. and {Koester}, D. and {Napiwotzki}, R. and {Christlieb}, N. and {Reimers}, D.},
        title = "{High-resolution UVES/VLT spectra of white dwarfs observed for the ESO SN Ia progenitor survey. II. DB and DBA stars}",
      journal = {\aap},
         year = 2007,
        month = aug,
       volume = {470},
       number = {3},
        pages = {1079-1088},
          doi = {10.1051/0004-6361:20077285},
       adsurl = {https://ui.adsabs.harvard.edu/abs/2007A&A...470.1079V}
}

@ARTICLE{petitclerc05,
       author = {{Petitclerc}, N. and {Wesemael}, F. and {Kruk}, J.~W. and {Chayer}, P. and {Bill{\`e}res}, M.},
        title = "{FUSE Observations of DB White Dwarfs}",
      journal = {\apj},
         year = 2005,
        month = may,
       volume = {624},
       number = {1},
        pages = {317-330},
          doi = {10.1086/428750},
       adsurl = {https://ui.adsabs.harvard.edu/abs/2005ApJ...624..317P}
}

\noindent


\end{document}